\documentstyle[12pt]{article}
\def\hbeta{{{\beta} \over {2}}}

\begin{document}

\begin{titlepage}
\rightline{CERN-TH/95-130}

\noindent
\begin{center}
{\Large{\bf Invisible Events with \\
Radiative Photons  at LEP } }
\end{center}
\vspace{1cm}
\bigskip

\begin{center}
{\large G.~Montagna~$^a$,
O.~Nicrosini$^{b}~\footnote{\footnotesize On leave from INFN, Sezione
di Pavia, Italy.}$, F.~Piccinini~$^c$ \\
and L.~Trentadue~$^b~\footnote{\footnotesize On leave from
Dipartimento di
Fisica, Universit\`a  di Parma, Parma, Italy.}~\footnote{\footnotesize
 INFN, Gruppo Collegato di Parma, Sezione di Milano, Milan,
Italy. }$} \\
\end{center}

\medskip\noindent
$^a$ INFN, Sezione di Pavia, Italy \\
\noindent
$^b$ CERN, TH Division, Geneva, Switzerland \\
\noindent
$^c$ INFN, Sezione di Pavia, Italy \\

\vspace{1cm}
\bigskip
\begin{center}
{\bf Abstract} \\
\end{center}
{\small A study of the radiative neutrino counting
reaction $e^+ e^- \to \nu {\bar \nu} \gamma$
at LEP1 and LEP2 energies is presented.
An approximate expression for the spectrum of the observed
photon is derived within the framework
of the $p_t$-dependent structure function approach.
This is compared with an exact expression and found in agreement within
the foreseen experimental accuracy. This model describing single-photon
radiation can be applied to the more general case of initial-state
single-photon emission accompanying invisible final-state events.
Higher-order QED corrections
due to undetected initial-state radiation are also included.
The implementation in a Monte Carlo event generator is briefly
described. }

\vskip 12pt \noindent
e-mail: \\
montagna@pv.infn.it, nicrosini@vxcern.cern.ch, \\
piccinini@pv.infn.it, trenta@vxcern.cern.ch

\vfil
\leftline{CERN-TH/95-130}
\leftline{May 1995}
\end{titlepage}

\noindent
\section{Introduction}
\vskip 10pt

The radiative neutrino-counting reaction and, more in general, the
single-photon reactions, provide a very useful tool for measuring the
number of light neutrinos and eventually detecting
new-physics signals at LEP~\cite{yrluca}.

The lowest-order spectrum for single-photon production within the
Standard Model is known~\cite{bbmmn}.
For energies around the $Z^0$ peak this exact calculation
agrees within about 1\% with two different
approximations~\cite{pia,piah}.

Concerning higher-order QED corrections, two different calculations
and the corresponding codes do exist~\cite{piah,mmm}.
Both are based on the QED structure
function approach~\cite{qed}.
In the first one~\cite{piah}, QED corrections are implemented
on the tree-level spectrum in the so-called Point Interaction
Approximation (PIA) and the corresponding program
is a semi-analytical code. In the second one~\cite{mmm},
 $\cal O(\alpha)$
plus higher-order QED corrections are included for the
$Z^0$-exchange only and
the contribution of the $W$-exchange
diagrams is added at the tree level. This
formulation is implemented in a Monte Carlo generator of
unweighted events. However, as is known, the treatment of the QED
radiative corrections around the $Z^0$ peak can become inadequate as
the energy increases. A proper treatment of the hard photonic radiation
has to be provided.

The purpose of this work is twofold: \\
i) to improve the approximations existing
in the literature for the study of the reaction at LEP1 energies,
in order to get reliable results for radiative
neutrino counting as well as for single-photon events at LEP2;
to provide a new result on the treatment of the hard-photon spectrum; \\
ii) to establish a general strategy for developing a single-photon
library, inclusive of all the standard and non-standard signals.
This last point is motivated by the
relevance of the experimental studies
 of single-photon radiation  accompanying invisible events
at LEP for the search of possible new-physics effects.

The outline of the paper is as follows. In Sect.~2 the
exact and approximate calculations for
the spectrum of the radiative neutrino counting reaction
at lowest order in the Standard Model are reviewed.
A quantitative study of the approximations at LEP2 versus LEP1
energies will be performed and an
approximate expression for the photonic spectrum will be presented.
In Sect.~3 it  will be shown how to include in the formulation
higher-order initial-state QED corrections, and our
results will be compared with existing ones.
Some single-photon distributions of
experimental interest will be shown and commented in Sect.~4.
Finally, conclusions and possible perspectives
on the subject are given in Sect.~5.

\vskip 15pt
\noindent
\section{Lowest-Order Calculations}
\vskip 10pt

The production of neutrino--antineutrino pairs and of
a single photon is
described within the Standard Model of the electroweak interactions by
$s$- and $t$-channel diagrams involving $Z^0$ and
$W^{\pm}$ exchange~\cite{yrluca}. $W$ contributions occur only for the
production of
 electron neutrinos and they can be
distinguished in diagrams with single-$W$
exchange and diagrams with the exchange of two $W$'s,
the latter involving
the non-Abelian trilinear coupling $WW\gamma$. On the basis of the
above diagrams, the differential cross section in the photon energy
$E_{\gamma}$ and angle $\cos{\vartheta}_{\gamma}$ has been
calculated exactly~\cite{bbmmn} and its expression
contains the $Z^0$ and $W$ squared amplitudes and
the $W$--$Z^0$ interference.
What is important to stress about this
calculation is that it is performed keeping a finite value for the mass
 of the $W$ boson and, therefore,
it applies without restrictions on the
energy range.

However, for the energies around the $Z^0$ peak, two approximations
are known in the literature and they both reproduce, with an
accuracy of the order of 1\%,
radiative neutrino counting experiments at LEP1. The two
approximations are:

\begin{itemize}

\item The Point Interaction (or Contact) Approximation~\cite{pia} (PIA).
It consists in neglecting the diagram with exchange of two $W$'s
and taking the limit $M_W \to \infty$ in the calculation.

\item The Point Interaction Approximation with Angular
Radiator~\cite{piah} (PIAH).
It uses  the QED $p_t$-dependent structure functions,
i.e. the method containing also the transverse degrees of freedom of
the emitted radiation~\cite{pt}. It consists in dressing the
cross section for producing a $\nu \bar \nu$ pair (without photons)
with an angle-dependent radiator. The photon spectrum is given in this
approximation by the following factorized expression:
\begin{eqnarray}
{{d \sigma} \over {d x_{\gamma} d c_{\gamma}} } =
\sigma_0^{PIA} ((1 \, - \, x_{\gamma}) \, s)
H^{(\alpha)} (x_{\gamma}, c_{\gamma}; s) .
\end{eqnarray}
$\sigma_0^{PIA}$ is the cross section for $e^+ e^- \to \nu \bar \nu$
as evaluated within the PIA approximation
at the energy scale reduced by photon emission;
$H^{(\alpha)}(x_{\gamma}, c_{\gamma}; s)$ is the
angle-dependent radiator which describes the probability of radiating
a photon with a given energy fraction
$x_{\gamma} = E_{\gamma} / E_b$ at the
angle $\vartheta_{\gamma}$ ($c_{\gamma} \equiv
\cos \vartheta_{\gamma}$) . Its
expression, derived from the ${\cal O}(\alpha)$ $p_t$-dependent
structure function based
on the $electron \to electron + photon$ splitting function,
reads as~\cite{piah}
\begin{eqnarray}
H^{(\alpha)}(x_{\gamma},p_t;s) =
{\alpha \over {2\pi}} \,P(1-x_{\gamma})  \frac{1}{\pi}
\Big[\frac{1}{p_{te^+}^2+x_\gamma^2m^2}+\frac{1}{p_{te^-}^2
+x_\gamma^2m^2}\Big]+ {\cal{O}}
(\frac {p_t^2} {E_{\gamma}^2} ) , \nonumber
\end{eqnarray}
where $P(1-x)$ is the Altarelli--Parisi splitting function and
$p_{te^{-,+}}$ represent the electron and positron transverse momenta,
respectively. In terms of the photon angle $c_\gamma$, one
obtains~\cite{piah}
\begin{eqnarray}
H^{(\alpha)}(x_{\gamma},c_{\gamma};s) = {\alpha \over \pi} \, { {1 + (1 -
x_{\gamma})^2} \over
{x_{\gamma}} } \, {1 \over {1 + 4 m^2 / s - c_{\gamma}^2} } .
\label{eq:has}
\end{eqnarray}
This expression contains the leading infrared and collinear
singularities
represented by the $1 / x_{\gamma}$ term and by the
angular factor in~(\ref{eq:has}), respectively.
This specific expression can, however, be
properly modified to include less dominant (next-to-leading)
contributions, as will be  shown in the following (see~\cite{piah}).
\end{itemize}

Figure~1 shows the integrated radiative neutrino counting cross section
with the minimum photon
tagging angle $\vartheta_{\gamma}^{min} = 20^\circ$ and
the minimum photon energy $E_{\gamma}^{min}= 1$~GeV. In this plot, the
solid line is the result obtained by
integrating the exact photon spectrum
of Ref.~\cite{bbmmn}, the dotted and dashed lines correspond to the
PIA~\cite{pia} and PIAH~\cite{piah} approximations,
respectively. As can be
seen, the two approximations are in good agreement with the
exact calculation around the $Z^0$ peak,
but they start to differ as the energy goes above about
120-130~GeV and the difference is well visible at
higher LEP2 energies.
Figure~2 is a blow-up of Fig.~1 at LEP1 (Fig.~2a) and
LEP2 (Fig.~2b) energies respectively.
The two approximations agree with the exact calculation
within $\simeq 1 \%$
at LEP1 (where they essentially provide the same integrated
cross section with a hard photon tail above the peak
due to the emission of a $Z^0$-return photon of energy
$E_{\gamma} \simeq (1 - M_Z^2 / s ) \sqrt{s} / 2$),
whereas at LEP2 they are in net disagreement with the
full spectrum. In particular, they
provide an increase in cross section with
energy, as a consequence of the absence of finite $W$-mass
effects needed at LEP2 energies.
Furthermore, the larger deviation of the PIAH
approximation is due, in particular, to the specific expression
 employed
for the angular radiator as in eq.~(\ref{eq:has}).

However, hard photons, strongly
suppressed around the $Z^0$ peak, are important at LEP2 and
they have to be included. Within the
framework of the
$p_t$-dependent structure function approach, by virtue of its
generality, a simple approximate expression can be achieved,
which constitutes one of the
results of the present paper. This is of some importance for
experiments
at LEP2 because, if a proper expression for handling the photon
spectrum is found, it can be used as a basic tool to develop a
library of single-photon events, including standard and non-standard
(in particular SUSY) processes. The approximate expression can be
implemented with the two following steps:
\begin{itemize}

\item to replace the cross section $e^+ e^- \to \nu \bar \nu$
in PIA approximation
 ($\sigma_0^{PIA}$)
with the corresponding exact cross section in the Standard Model
($\sigma_0^{exact}$), i.e. for
$e^+ e^- \to (Z,W) \to \nu \bar \nu$, thus including the necessary
$W$-mass effects~\footnote{\footnotesize
The exact invisible
cross section was
computed by means of {\tt SCHOONSCHIP}~\cite{schoon}.};

\item to adopt, instead of the form
in eq.~(\ref{eq:has}), the more accurate expression for the
angular radiator in such a
way that also a part of next-to-leading contributions
are included. This properly modified radiator has been already
derived~\cite{piah} and reads:
\begin{eqnarray}
H^{(\alpha)}_{SH} = {\alpha \over {2 \pi} } \,
{1 \over x_{\gamma} } \, \left[ 2 \, { {1 + (1 - x_{\gamma})^2} \over
{1 + 4 m^2 / s - c_{\gamma}^2} } - x_{\gamma}^2 \right] .
\label{eq:hsh}
\end{eqnarray}

\end{itemize}
The latter expression differs from the radiator of
eq.~(\ref{eq:has})
by less than 1\% around the $Z^0$ peak,
 but it is more appropriate to describe the emission of hard
photons (of energy fraction $x_{\gamma} \simeq 1$) at large angles.

Therefore the photonic spectrum of the
reaction $e^+ e^- \to \nu \bar \nu \gamma$ can be written
in factorized form as follows:
\begin{eqnarray}
{{d \sigma} \over {d x_{\gamma} d c_{\gamma}} } =
\sigma_0^{exact} ((1 \, - \, x_{\gamma}) \, s)
H^{(\alpha)}_{SH} (x_{\gamma}, c_{\gamma}; s) .
\label{eq:app2}
\end{eqnarray}

Let us comment about the physical meaning of the above equation.
Given the specific bare kernel $e^+ e^- \to \nu \bar \nu$,
as well as for any other kernel corresponding to any
given process of the type $e^+ e^- \to invisible$ objects,
 dressing it
with the angular radiator $H^{(\alpha)}_{SH}$ amounts to attaching
a photon line on the external charged legs; in this way,
the ``universal'', factorized form of the photonic radiation will be
included,
 comprehensive of
some of the next-to-leading effects. Therefore, in this approximation
the $W$ diagram with the photon emitted by the internal propagator is not
reproduced.

However,
by virtue of the generality of the method, if
 the cross section for an $e^+ e^-$ scattering
yielding an invisible final state
 is known, an approximation of the cross
section for the corresponding radiative process
can be obtained in a straightforward way.

Figure~3 shows a comparison of the exact spectrum with the
approximate results obtained
 using the two different angle-dependent radiators
of~(\ref{eq:has}) and (\ref{eq:hsh}), respectively. The solid line
corresponds to the exact spectrum; the dotted and dashed lines are the
result derived from convoluting the exact
bare cross section with
the angular radiators of eq.~(\ref{eq:has})
and eq.~(\ref{eq:hsh}), respectively.

As can be clearly seen, whereas the
approximation based on the
radiator of eq.~(\ref{eq:has}) differs by several per cent from the
exact result, using the ``modified'' radiator of
eq.~(\ref{eq:hsh}) improves the agreement at the
level of a few per cent.
To be more precise, for $150 \leq \sqrt{s} \leq 175$
the relative difference between the exact results and the approximation
of eq.~(\ref{eq:app2}) is of the order of 1--2$\%$ and becomes
of the order of 3--4$\%$ only at about 200~GeV.
These deviations, depending
on the centre-of-mass energy, can be seen as a measure of the
non-leading effects not included in the angular radiator approximation.
However, for the experimental precision foreseen at LEP2, a theoretical
 error of the order of a few per cent is more than adequate.

\vskip 15pt
\section{Higher-Order QED Corrections}
\vskip 10pt

To match the experimental precision expected at LEP2, the corrections due to
multiphoton soft emission and to the radiation of hard photons
 emitted in the very forward direction (and hence lost in the beam pipe)
by the colliding beams must be included. The large QED corrections introduced
by the undetected initial-state radiation can be successfully described
within the framework of the QED structure function
approach. This method, as recalled in the Introduction,
is indeed  employed in
both the formulations
developed for radiative neutrino counting at LEP1~\cite{piah,mmm}.
 However, both the approaches have
their drawback and therefore need to be updated for
experiments at LEP2. A simple solution, which represents an improvement
of both the existing approaches and has an overall accuracy of
about $1\%$, can be achieved by convoluting the exact spectrum of
Ref.~\cite{bbmmn}
 with electron and positron structure functions. The QED-corrected
cross section can then be written, in analogy with QCD
factorization, as a convolution of the following
form~\cite{prd}:
\begin{eqnarray}
\sigma (s) = \int d x_1 \, d x_2 \, d E_{\gamma} \, d c_{\gamma}
\, D(x_1,s) D(x_2,s) {{d \sigma} \over {d E_{\gamma} d c_{\gamma}}} ,
\label{eq:master}
\end{eqnarray}
where $d \sigma / d E_{\gamma} d c_{\gamma}$
is the exact spectrum of $e^+ e^- \to \nu {\bar \nu}
\gamma$~\cite{bbmmn}, the photon variables being referred to the
centre-of-mass frame after initial-state radiation, and $D(x,s)$ is
the electron (positron) structure function.
Its expression, as obtained by solving the Lipatov--Altarelli--Parisi
evolution equation in the non-singlet approximation, is given
by~\cite{qed}:
\begin{eqnarray}
D(x,s)&=&\, { {\exp \left\{ \frac{1}{2}\beta \,\left( \frac{3}{4} -\gamma_E
\right) \right\} }
\over {\Gamma \left( 1 + \frac{1}{2} \beta \right) } } \,
\hbeta (1 - x)^{\hbeta - 1} - {{\beta} \over 4} (1+x) \\ \nonumber
&+& {1 \over {32}} \beta^2 \left[ -4 (1+x) \ln(1-x) + 3 (1+x) \ln x -
4 {{\ln x} \over {1-x}} - 5 - x \right] ,
\end{eqnarray}
with
\begin{eqnarray}
\beta = 2\,{\alpha \over \pi}\,(L-1) ,
\end{eqnarray}
\noindent
where $L= \ln \left( {s / {m^2}} \right)$ is the collinear logarithm,
$\gamma_E$ is the Euler constant and $\Gamma(z)$ is the gamma function.
The first exponentiated Gribov--Lipatov term
describes multiphoton soft emission,
the second and third ones hard bremsstrahlung
in the collinear approximation. Further non-leading QED corrections are
included as a $K$-factor, computed from the exact $O(\alpha)$ result
of~Ref.~\cite{ignak}. The relevant weak corrections are taken into
account, for the $Z^0$-exchange contribution,
in the form of Improved Born Approximation, according to the
general recipe of Ref.~\cite{weak}.

The master formula~(\ref{eq:master}) is implemented in a Monte Carlo
event generator~\cite{cpcnu}~\footnote{\footnotesize
For a first attempt
to develop a neutrino counting Monte Carlo based on $p_t$-dependent
structure functions, see~Ref.~\cite{moe}. }.
Figure~4a shows the QED-corrected
cross section as a function of the centre-of-mass
energy at LEP2. The dash-dotted
line represents the exact cross section at lowest order; the solid
line is the result corresponding to eq.~(\ref{eq:master}) (i.e. in the case
of convolution of the full spectrum), and the dotted line, reported
for the sake of comparison, shows the results obtained by simulating the
approach of Ref.~\cite{mmm}, namely correcting the $Z^0$ contribution
only, and adding to this result the $W$-exchange diagrams at tree level.
Two considerations are in order here. First, the QED-corrected cross section
is higher than the Born one as a consequence of the $Z^0$ radiative
return:
the effect is to enhance the Born result by a factor of about 1.3.
Secondly, the convolution of the full spectrum
is in good agreement (within 1\%) with the approach of Ref.~\cite{mmm}
 because the QED-corrected cross section is largely dominated by
the $Z^0$ radiative return, and the tree-level contribution of
$W$ diagrams (and $W$--$Z$ interference) is almost flat over the
full energy range spanning from LEP1 to LEP2~(Fig.~4b). Of course, a
more general (beyond the Standard Model) single-photon spectrum,
including initial-state QED corrections, can be simply obtained. This
can be achieved by substituting, in eq.~(5), the exact proper
generalized spectrum (or, if sufficient, its approximation according to
eq.~(4)) for $d \sigma / d E_\gamma d c_\gamma$ within the Standard
Model.

\vskip 15pt
\section{Simulation of Photon Distributions}
\vskip 10pt

In this section, some technical details about
the Monte Carlo event generator developed on the basis of the
formulation presented in the previous section will be sketched;
a complete and
exhaustive account of the code will be given elsewhere~\cite{cpcnu}.
A sample of realistic experimental distributions
obtained by analysing the single-photon events generated by the
Monte Carlo code will also be shown and briefly  discussed.

Since the integrand is a superposition of different peaking
behaviours, the standard importance-sampling
technique, within the multichannel approach, was employed
to develop an efficient event generator.
The QED-corrected
cross section is written as the sum of four different channels
with the following features: (1) the radiative variables
$x_{1,2}$ are generated according to the soft exponentiated
part of the structure function,
 and the photon energy is then generated according to
the dependence of
the $Z^0$ propagator on $E_\gamma$; (2)~$x_{1,2}$ as above and
$E_\gamma$ generated according to the infrared-like behaviour
$1 / E_\gamma$; (3)~$E_\gamma$ flat, $x_1$ as the soft part of the
structure function
and $x_2$ according to the dependence of the $Z^0$ propagator
on the reduced centre-of-mass energy
${\hat s} = x_1 x_2 s$; (4) the same as (3)
with the interchange of $x_1 \leftrightarrow x_2$. Moreover, to
take care of the Lorentz boost caused by the emission of
unbalanced electron and positron radiation, the photon phase-space
variables are first generated in the centre-of-mass
frame and then boosted
to the laboratory frame by the boost given by the Lorentz factor
$\beta_L = (x_1 - x_2) / (x_1 + x_2)$ valid for collinear
kinematics.

Figures 5, 6 and 7 show the photon energy distribution,
the photon angular distribution and the $p_t^{\gamma} / E_b$ distribution
at a fixed LEP1 ($E_b = 48$~GeV) and LEP2 ($E_b = 88$~GeV) beam
energies,
respectively, assuming an experimental
apparatus with $E_{\gamma} = 1$~GeV and
$\vartheta_{\gamma}^{min} = 20^\circ$.
The solid histograms correspond to the lowest-order
approximation, the dashed ones include the effect
of initial-state radiation. In order to consistently compare the
tree-level and QED-corrected distributions, the numbers of events
integrated in the two cases are proportional to the corresponding
integrated cross sections.
Concerning the
photon energy distribution~(Fig.~5), two peaks are well
visible both at LEP1 (Fig.~5a) and
LEP2 (Fig.~5b) energies: the higher
one is located at the energy value of about
$(1- M_Z^2 / s) \sqrt{s} / 2$, the
 lower one is due to $1 / E_{\gamma}$ peaking behaviour.
As can be seen, the
main modifications introduced by initial-state radiation
are to reduce the higher peak and to enhance the lower one, as expected
by typical convolution effects,
thus modifying the two-bump
asymmetrical shape of the Born distribution
into a more symmetrical one. The photon angular
distribution (Fig.~6) has the
same symmetrical shape
(peaked in the forward direction) at LEP1 (Fig.~6a) and
LEP2 (Fig.~6b) energies,
the only difference
being that at LEP2 the QED-corrected distribution is higher than the
Born one because of the $Z^0$ radiative return. At last,
comparing the $p_t^{\gamma} / E_b$ distributions at LEP1 (Fig.~7a)
and LEP2 (Fig.~7b) energies the emission of mostly
soft photons at LEP1 and of harder ones at larger angles at LEP2 can be clearly
recognized.

\vskip 15pt
\section{Conclusions}
\vskip 10pt

In this paper a study of the
neutrino counting reaction $e^+ e^- \to \nu {\bar \nu} \gamma$ at LEP2
energies has been presented. Within the
$p_t$-dependent structure function approach,
an approximate expression for the photon energy and angle spectrum
has been derived.
The accuracy of the approximation, as
estimated by comparison with the exact result, is of the order of
$1$--$2\%$ near the $W$-pair threshold and grows larger (up to
about 3--4\%) only at about 200~GeV. By virtue of its generality,
the approximation can be used as a first attack strategy
 to dress standard as well as non-standard (e.g. SUSY)
processes giving rise to an
invisible final state in order to get
an approximate formula for the spectrum of the corresponding radiative
single-photon events. Therefore, on the basis of these considerations,
a library of single-photon processes should be feasible for
 experiments at LEP2. A further investigation on the tree-level
generalized spectra would be of course required for processes whose
form is less simple than the one shown in eq.~(4).

Concerning higher-order
QED corrections, a formulation based
on structure function
convolution of the exact photon spectrum is  presented,
 which can be estimated to be accurate at the 1\% level
and agrees with
the approximate approach of Ref.~\cite{mmm} within about 1\%.
This is implemented in a Monte Carlo event generator~\cite{cpcnu},
which also can be
easily generalized to describe single-photon radiative
processes accompanying invisible events at LEP.

\vskip 15pt
\leftline{\bf Acknowledgements}
\vskip 10pt \noindent
Useful discussions with B.~Mele and G.~W.~Wilson
 are gratefully acknowledged.

\vfill\eject

\leftline{\large \bf Figure Captions}
\vskip 30pt
\noindent
Figure 1. The lowest-order
integrated cross section of $e^+ e^- \to \nu {\bar \nu}
\gamma$ for
$E_{\gamma} = 1$~GeV and $\vartheta_{\gamma}^{min} = 20^\circ$
as a function of the
centre-of-mass energy. Solid line: exact spectrum; dotted line: PIA;
dashed line: PIA with angular radiator.

\vskip 8pt\noindent
Figure 2. Blow-up of Fig.~1 at LEP1 (Fig. 2a) and LEP2 (Fig. 2b) energies.

\vskip 8pt\noindent
Figure 3. The lowest-order integrated cross section at LEP2.
Solid line: exact spectrum; dotted line: ``soft'' angular radiator of
eq.~(2);
dashed line: ``modified'' angular radiator of eq.~(3).

\vskip 8pt\noindent
Figure 4. The QED-corrected integrated cross section at LEP2 (Fig.~4a).
Solid line: convolution of the
exact spectrum; dotted line: convolution of the $Z^0$ contribution with
$W$ at tree level; dash-dotted line: exact Born spectrum.
The $W$ and $W$--$Z^0$ interference contributions as a function
of the centre-of-mass energy (Fig.~4b).

\vskip 8pt\noindent
Figure 5.  The photon energy distributions at LEP1 (Fig.~5a) and LEP2
(Fig.~5b) energies.
Solid histogram: Born; dashed histogram: QED-corrected.

\vskip 8pt\noindent
Figure 6. The same as Fig. 5 for the photon angular distribution.

\vskip 8pt\noindent
Figure 7. The same as Fig. 5 for the $p_t^{\gamma} / E_b$ distribution.
\vfil\eject


\vskip 24pt
\begin{thebibliography}{99}

\bibitem{yrluca}{L.~Trentadue et al., {\it Neutrino
Counting}, in {\it Z Physics at LEP1},
G.~Altarelli, R.~Kleiss and C.~Verzegnassi, eds., CERN 89--08, Vol.~1
(1989) p.~129, and references therein.}

\bibitem{bbmmn}{F.~A.~Berends et al., Nucl.~Phys.~B301 (1988) 583.}

\bibitem{pia}{K.~J.~F.~Gaemers, R.~Gastmans and F.~M.~Renard,
Phys.~Rev.~ D19 (1979) 1605; \\
G.~Barbiellini, B.~Richter and J.~L.~Siegrist, Phys.~Lett.
B106 (1981) 414.}

\bibitem{piah}{O.~Nicrosini and L.~Trentadue, Nucl.~Phys.
B318 (1989) 1.}

\bibitem{mmm}{R.~Miquel, C.~Mana and M.~Martinez,
Z.~Phys.~C48 (1990) 309; also in {\it QED Structure
Functions}, G.~Bonvicini, ed., AIP Conf. Proc. No.~201
(AIP, New York, 1990), p.~395. }

\bibitem{qed} {E.~A.~Kuraev and V.~S.~Fadin, Sov.~J.~Nucl.~Phys.
41 (1985) 466; \\
G.~Altarelli and G.~Martinelli, {\it Physics at LEP},
CERN Report 86--02, J.~Ellis and R.~Peccei, eds. (Geneva, 1986);
see also: \\
O.~Nicrosini and L.~Trentadue, Phys.~Lett.
B196 (1987) 551;
Z.~Phys.~C39 (1988) 479. \\
For a review see also: \\
O.~Nicrosini and L.~Trentadue,
in {\it Radiative Corrections for $e^+ e^-$ Collisions},
J.~H.~K\"uhn, ed. (Springer, Berlin, 1989), p.~25; in {\it QED Structure
Functions}, G.~Bonvicini, ed., AIP Conf. Proc. No.~201
(AIP, New York, 1990), p.~12; O.~Nicrosini, ibid., p.~73. }

\bibitem{pt}{O.~Nicrosini and L.~Trentadue, Phys.~Lett. B231 (1989)
487, and references therein. \\
For a review see also: \\
O.~Nicrosini and L.~Trentadue,
in {\it Radiative Corrections for $e^+ e^-$ Collisions},
J.~H.~K\"uhn, ed. (Springer, Berlin, 1989) p.~25; in {\it QED Structure
Functions}, G.~Bonvicini, ed., AIP Conf. Proc. No.~201
(AIP, New York, 1990), p.~12; O.~Nicrosini, ibid.,  p.~73. }

\bibitem{schoon} {{\tt SCHOONSCHIP}, {\it A Program for Symbol Handling} by
M.~Veltman, see H.~Strubbe, Comput. Phys. Commun. 8 (1974) 1. }


\bibitem{prd}{G.~Montagna, O.~Nicrosini and F.~Piccinini,
Phys.~Rev.~D48 (1993) 1021, and references therein; \\
M.~Cacciari et al., Phys. Lett. B268 (1991) 441. }

\bibitem{ignak} {M.~Igarashi and N.~Nakazawa, Nucl.~Phys.~B288 (1987)
301;  H. Veltman,  Nucl. Phys. B312 (1989) 1. }

\bibitem{weak} {M.~Consoli, W.~Hollik and F.~Jegerlehner, {\it
Electroweak radiative corrections for $Z$ physics},
in {\it Z Physics at LEP1},
G.~Altarelli, R.~Kleiss and C.~Verzegnassi, eds., CERN 89--08, Vol.~1
(1989), p.~7, and references therein.}

\bibitem{moe} {G.~Bonvicini and L.~Trentadue, Nucl. Phys. B323 (1989)
253; also in {\it QED Structure
Functions}, G.~Bonvicini, ed., AIP Conf. Proc. No.~201
(AIP, New York, 1990), p.~216. }

\bibitem{cpcnu}{G.~Montagna, O.~Nicrosini and F.~Piccinini, in
preparation (to be submitted to Comput. Phys. Commun.).}

\end{thebibliography}
\end{document}